# Chemical Physics of Controlled Wettability and Super Surfaces

Carolina Brito[1], Hans-Jürgen Butt[2], Alberto Giacomello[3]
1) Instituto de Física, Universidade Federal do Rio Grande do Sul, Caixa Postal 15051, CEP 91501-970, Porto Alegre, Rio Grande do Sul, Brazil
2) Max Planck Institute for Polymer Research, Mainz, Germany
3) Dipartimento di Ingegneria Meccanica e Aerospaziale, Sapienza Università di Roma, 00184, Rome, Italy

## INTRODUCTION

Wetting phenomena are widespread in both natural and technological contexts. Despite the well-established nature of this scientific field and our extensive knowledge of its underlying principles, wetting remains a dynamic and vibrant area of study. It continues to pose fundamental questions while offering innovative avenues for controlling these phenomena to develop novel applications.

By tailoring the wetting properties of surfaces, researchers and engineers can design materials with specific functionalities, such as self-cleaning surfaces, anti-fog coatings, and enhanced slipperiness. Recent years have witnessed significant advancements in wetting research, owing to the exquisite control achieved in surface topography and chemistry and to the development of novel experimental techniques. Additionally, simulations and theory have played a crucial role in these advancements. They provid the fundamental knowledge and quantitative tools to control wettability and design surfaces with enhanced properties.

Given these recent breakthroughs, this special collection *Chemical Physics of Controlled Wettability and Super Surfaces* becomes particularly timely and significant. It serves as a platform to showcase some of the latest developments in the field of wetting. It highlights the exciting progress and potential applications in controlling wetting properties that are enabled by the synergy between theory, simulations, and experiments.

## SUMMARY OF AREAS COVERED

This special collection covers 19 articles whose topics span the fundamental scientific questions of the field and explore various potential applications arising from the control of wetting properties. These investigations employ a blend of preparatory, experimental, numerical methodologies, and theoretical models to address diverse questions concerning the manipulation of wetting properties.

This Collection demonstrates that several fundamental aspects of wetting are still open to investigation. Reference [19] explores the origin of hydrophobic solvation phenomena via theory, classical density functional theory, and atomistic simulations. By considering strongly hydrophobic solutes of varying radii, the authors argue that density depletion and enhanced fluctuations can be related to the critical drying surface phase transition that occurs at bulk liquid–vapor coexistence for a planar substrate. In the context of machine learning applied to classical density functional theory, reference [1] introduces a Bayesian inference approach to reconstruct the external potential acting on a many-particle system. This approach is tested on a simple one-dimensional case, which illustrates the potential of such approaches in applications in which the wettability of surfaces is of interest. In reference [8], a phase field simulation method is developed and used to systematically study liquid filling on grooved surfaces. This approach allows to tune the nature and range of the liquid-solid interactions, capturing different wetting regimes (complete, partial, and pseudo-partial wetting states). In reference [7], a classical nucleation theory is developed to predict the effects of dissolved gasses, such as nitrogen and carbon dioxide, on liquid behavior in cylindrical nanopores.

A fundamental thermodynamic question is addressed in reference [5]. They address the relation between vapor of a liquid, a finite contact angle and a possible wetting transition at saturation vapor pressure. In order to have defined contact angles for liquids at saturated vapor pressures the corresponding adsorption isotherms needs to have a finite value at saturated conditions. They demonstrate that for water on silica this is the case.

Ink spreading and pattern formation are influenced by substrate wettability. Printed lines are unstable on surfaces with low wettability. To overcome the challenge of printing with precision on hydrophobic surfaces, the authors have conducted a combination of simulation and experimental work in reference [2].

Papers [4] and [10] exemplify the manipulation of wetting properties for the purpose of atmospheric water harvesting. In [4] the authors realized different substrates capable of collecting water from the atmosphere at low subcooling conditions, exploring the role of contact angle and contact angle hysteresis. Although hydrophilic and hydrophobic coatings were found to have a similar water collection efficiency, on the former the droplet departure was much faster. In reference [10] dynamic simulations were used to investigate the flow of water within hydrophobic nanocones decorated with hydrophilic rings. Their study encompasses an examination of the spatial arrangement of these hydrophobic rings within an individual nanocone and explores the synergistic effects of multiple nanocones to enhance water flux across these structures.

Reference [15] explores the wetting properties to design anti-fouling substrates. The authors study a very interesting phenomenon where crystalline structures, forming from evaporating saline water droplets, spontaneously detach – or self-eject – from superhydrophobic materials. This phenomenon is better understood when the water contains a specific concentration of salt. In this study, they investigate how the presence of certain contaminants affects the self-ejection. They show that certain contaminants can facilitate ejection even under conditions where it was previously not observed in pure sodium chloride solutions.

Droplet mobility is a relevant aspect in applications where minimizing adhesion and friction is essential to maintain the droplet in a non-wetted state on the substrate. Several papers in this collection delve into aspects related to droplet mobility to provide a deeper understanding of these phenomena. In reference [6], the authors conducted experiments on tilted surfaces to compare the velocity of two distinct types of non-wetting water droplets. The first type, referred to as 'pearls,' exhibits hydrophobic properties originating when a water droplet comes into contact with a rough substrate. The second type, termed 'marbles', achieves a non-wetting state by applying a hydrophobic powder to the droplet's surface, effectively isolating the water from the substrate. Reference [12] uses simulations of a droplet's dynamic behavior when subjected to a wetting gradient and external forces. These simulations enable a comprehensive characterization of the droplet's dynamics and provide insights into the various mechanisms governing droplet mobility, particularly in the context of competing forces. Reference [3] reports an experimental study complemented by simulations to systematically study and describe the dynamics of evaporating drops on surfaces with micro-patterns of triangular posts.

In reference [13], the impact of substrate fractality on the wetting behavior of droplets is investigated through a combination of a simple theoretical model and simulations employing the 3-spin Potts model for surfaces featuring hierarchical structures. This study provides insights into some open questions, including the relationship between the fractal dimension of the substrate and the contact angle of the droplet.

Reference [18] investigates numerically the pressure drop reduction in microchannels induced by the presence of liquid-infused walls. The pressure drop was found to be most prominently affected by the viscosity ratio between the two fluids. The phase field simulation methods also allowed the authors to determine the shape of the interface between the flowing liquid and lubricant.

Several papers address or apply surfaces with low contact angles hysteresis. Low contact angle hysteresis implies that contact lines or sessile drops move without much resistance. In the reference [16], the authors apply surfaces with low contact angle hysteresis to the evaporation of drops of salt solutions. Usually salt crystals are formed as soon as the salt concentration in the evaporating drop exceeds the saturation concentration. The contact line is pinned and stains are formed on the surface. The authors show that surfaces with low contact angle hysteresis suppress crystallization of salt. Evaporation proceeds at constant contact angle till all water has evaporated.

In [9] surfaces with low contact angles hysteresis are used to study the contact between two immiscible droplets. When two immiscible sessile droplets get into contact, two four-phase contact points are formed. They show that the four-phase contact point dynamics is faster on PDMS coated surfaces than on surfaces hydrophobized by silanes. In addition, the corresponding position-vs-time graph follows a different power law.

Two papers address the wetting dynamics of polymer brushes. These surfaces have attracted attention because they present low contact angle hysteresis. In [11], the authors analyze the transport of liquid at the rim of a hexadecane droplet on a hydrophobic polymer brush. They place the drop on the brush and measure the thickness of the brush around the drop interferometrically. Although hexadecane has a low vapor pressure, transport through the vapor phase occurs in parallel to lateral diffusion in the brush. The experimental measurements are well described by a gradient dynamics model.

The dynamic wetting of reactive brushes was further studied in [14]. They let water drops run down tilted surfaces coated with a brush with hydrolysable side groups. They observe a transition in the advancing contact angle at a defined drop velocity. From the transition velocity they can estimate the time constant for the hydrolysis.

Large contact angle hysteresis surfaces were the topic of a contribution in reference [17]. They analyzed the wetting of a porous surface formed by $TiO_2$ nanoparticles and stearic acid on copper. By monitoring the electrical current through the water droplet to the copper substrate, they verify that the water drops penetrate the porous layer to make direct contact with the copper. This penetration enhances the adhesion of the droplet to the film and helps to understand the contact angle hysteresis.

**CONCLUSIONS**

The articles featured in this special issue offer an insightful snapshot of the breadth within the realm of wettability and its diverse applications in crafting surfaces with unique properties. These papers shed light on various challenges tied to manipulating surface wetting, aimed at leveraging them for a range of functions and unraveling fundamental aspects underlying these phenomena. We hope that this compilation will captivate the interest of our community, sparking fresh perspectives to overcome obstacles and ignite innovative concepts for harnessing wetting principles in the creation of novel material classes.

**ACKNOWLEDGEMENTS**
We would like to thank the authors who participated in this special issue and the reviewers, whose dedicated work allowed the selection of the papers included here. We also acknowledge the efforts of *The Journal of Chemical Physics* editors Mark Ediger, Francesco Sciortino, and Carlos Vega, journal staff Olivia Zarzycki, and editor-in-chief, Tim Lian, for their assistance and efficiency during this process.

**REFERENCES**

[1] Antonio Malpica-Morales, Peter Yatsyshin, Miguel A. Durán-Olivencia, Serafim Kalliadasis; Physics-informed Bayesian inference of external potentials in classical density-functional theory. *J. Chem. Phys.* 14 September 2023; 159 (10): 104109. https://doi.org/10.1063/5.0146920

[2] Paria Naderi, Benjamin Raskin Sheuten, Alidad Amirfazli, Gerd Grau; Inkjet printing on hydrophobic surfaces: Controlled pattern formation using sequential drying. *J. Chem. Phys.* 14 July 2023; 159 (2): 024712. https://doi.org/10.1063/5.0149663

[3] Hsuan-Yi Peng, Bang-Yan Liu, Chi-Chun Lo, Li-Jen Chen, Ralf Seemann, Martin Brinkmann; De-wetting of evaporating drops on regular patterns of triangular posts. *J. Chem. Phys.* 14 July 2023; 159 (2): 024704. https://doi.org/10.1063/5.0151236

[4] Anthony Katselas, Isaac J. Gresham, Andrew R. J. Nelson, Chiara Neto; Exploring the water capture efficiency of covalently attached liquid-like surfaces. *J. Chem. Phys.* 7 June 2023; 158 (21): 214708. https://doi.org/10.1063/5.0146847

[5] Sepehr Saber, Nagarajan Narayanaswamy, C. A. Ward, Janet A. W. Elliott; Experimental examination of the phase transition of water on silica at 298 K. *J. Chem. Phys.* 28 May 2023; 158 (20): 204712. https://doi.org/10.1063/5.0145932

[6] Panlin Jin, Kexin Zhao, Zoé Blin, Malou Allais, Timothée Mouterde, David Quéré; When marbles challenge pearls. *J. Chem. Phys.* 28 May 2023; 158 (20): 204709. https://doi.org/10.1063/5.0150082

[7] Hikmat Binyaminov, Janet A. W. Elliott; Quantifying the effects of dissolved nitrogen and carbon dioxide on drying pressure of hydrophobic nanopores. *J. Chem. Phys.* 28 May 2023; 158 (20): 204710. https://doi.org/10.1063/5.0146952

[8] Fandi Oktasendra, Arben Jusufi, Andrew R. Konicek, Mohsen S. Yeganeh, Jack R. Panter, Halim Kusumaatmaja; Phase field simulation of liquid filling on grooved surfaces for complete, partial, and pseudo-partial wetting cases. *J. Chem. Phys.* 28 May 2023; 158 (20): 204501. https://doi.org/10.1063/5.0144886

[9] Peyman Rostami, Mohammad Ali Hormozi, Olaf Soltwedel, Reza Azizmalayeri, Regine von Klitzing, Günter K. Auernhammer; Dynamic wetting properties of PDMS pseudo-brushes: Four-phase contact point dynamics case. *J. Chem. Phys.* 15 May 2023; 158 (19): 194703. https://doi.org/10.1063/5.0142821

[10] Fernanda R.Leivas, Marcia C. Barbosa; Functionalized carbon nanocones performance in water harvesting. *J. Chem. Phys.* 15 May 2023; 158 (19): 194702. https://doi.org/10.1063/5.0142718

[11] Özlem Kap, Simon Hartmann, Harmen Hoek, Sissi de Beer, Igor Siretanu, Uwe Thiele, Frieder Mugele; Nonequilibrium configurations of swelling polymer brush layers induced by spreading drops of weakly volatile oil. *J. Chem. Phys.* 7 May 2023; 158 (17): 174903. https://doi.org/10.1063/5.0146779

[12] Leon Topp, Lena Haddick, Dominik Mählmann, Andreas Heuer; Wettability gradient-driven droplets with an applied external force. *J. Chem. Phys.* 7 May 2023; 158 (17): 174703. https://doi.org/10.1063/5.0146910

[13] Iara Patrícia da Silva Ramos, Cristina Gavazzoni, Davi Lazzari, Carolina Brito; Hierarchical structured surfaces enhance the contact angle of the hydrophobic (meta-stable) state. *J. Chem. Phys.* 21 April 2023; 158 (15): 154703. https://doi.org/10.1063/5.0146948

[14] Xiaomei Li, Krisada Auepattana-Aumrung, Hans-Jürgen Butt, Daniel Crespy, Rüdiger Berger; Fast-release kinetics of a pH-responsive polymer detected by dynamic contact angles. *J. Chem. Phys.* 14 April 2023; 158 (14): 144901. https://doi.org/10.1063/5.0142928

[15] Samantha A. McBride, John R. Lake, Kripa K. Varanasi; Self-ejection of salts and other foulants from superhydrophobic surfaces to enable sustainable anti-fouling. *J. Chem. Phys.* 7 April 2023; 158 (13): 134721. https://doi.org/10.1063/5.0142428

[16] Alex Jenkins, Gary G. Wells, Rodrigo Ledesma-Aguilar, Daniel Orejon, Steven Armstrong, Glen McHale; Suppression of crystallization in saline drop evaporation on pinning-free surfaces. *J. Chem. Phys.* 28 March 2023; 158 (12): 124708. https://doi.org/10.1063/5.0139448

[17] L. E. Helseth, M. M. Greve; Wetting of porous thin films exhibiting large contact angles. *J. Chem. Phys.* 7 March 2023; 158 (9): 094701. https://doi.org/10.1063/5.0138148

[18] Amirmohammad Rahimi, Arghavan Shahsavari, Hossein Pakzad, Ali Moosavi, Ali Nouri-Borujerdi; Laminar drag reduction ability of liquid-infused microchannels by considering different infused lubricants. *J. Chem. Phys.* 21 February 2023; 158 (7): 074702. https://doi.org/10.1063/5.0137100

[19] Mary K. Coe, Robert Evans, Nigel B. Wilding; Understanding the physics of hydrophobic solvation. *J. Chem. Phys.* 21 January 2023; 158 (3): 034508. https://doi.org/10.1063/5.0134060